\documentclass[12pt]{article}

\usepackage{amsfonts}
\usepackage{amsmath,amscd}

\numberwithin{equation}{section}
\renewcommand{\theequation}{\thesection.\arabic{equation}}

\def\appendix#1{\addtocounter{section}{1}\setcounter{equation}{0}
\renewcommand{\thesection}{\Alph{section}}
\section*{Appendix \thesection\protect\indent \parbox[t]{11.15cm}{#1}}
\addcontentsline{toc}{section}{Appendix \thesection\ \ \ #1}}

\begin{document}

\begin{titlepage}
\vfill
\begin{flushright}
\end{flushright}
\vfill
\begin{center}
   \baselineskip=16pt
  {\Large\bf Towards Cosmological Black Rings}
   \vskip 2cm
       Jan B. Gutowski$^1$, 
      and W. A. Sabra$^2$\\
 \vskip .6cm
      \begin{small}
      $^1$\textit{Department of Mathematics, King's College London.\\
      Strand, London WC2R 2LS\\United Kingdom \\
        E-mail: jan.gutowski@kcl.ac.uk}
        \end{small}\\*[.6cm]
  \begin{small}
      $^2$\textit{Centre for Advanced Mathematical Sciences and
        Physics Department, \\
        American University of Beirut, Lebanon \\
        E-mail: ws00@aub.edu.lb}
        \end{small}      
   \end{center}
\vfill
\begin{center}
\textbf{Abstract}
\end{center}
\begin{quote}
We classify all pseudo-supersymmetric near horizon geometries of extremal black holes in
five dimensional de-Sitter supergravity coupled to vector multiplets.
We find that there are three types of solution. The first type corresponds
to the near-horizon geometry of extremal de Sitter BMPV black holes,
and the spatial cross-section of the horizon is topologically $S^3$.
The other two solutions cannot be embedded into the minimal de-Sitter supergravity theory,
and correspond to near-horizon geometries for which the spatial cross-sections of the horizon 
are $T^3$ or $S^1 \times S^2$. 

\end{quote}
\end{titlepage}

\section{Introduction}

The discovery of five-dimensional black ring solutions
for which the spatial cross-sections of the event horizon have topology $S^1 \times S^2$ 
\cite{ring1, ring2, ring3, ring4, ring5, ring6}
has shown that there are many interesting black objects in higher dimensions.
In particular, the black hole uniqueness theorems formulated originally in four dimensions \cite{unique1, unique2,
unique3, unique4, unique5, unique6}
do not generalize to higher dimensions, though uniqueness theorems have been formulated 
for static solutions in higher dimensions \cite{gibbons1, rogatko}, and for solutions with additional rotational Killing vectors
\cite{extr3, extr4, extr5}. It is notable that all known black ring solutions are asymptotically flat. The status of 
black rings which are not asymptotically flat is unresolved. In particular,  asymptotically
anti-de-Sitter or de-Sitter black ring solutions with regular horizons have yet to be found,
though approximate solutions have been constructed in \cite{adsring1}. 

One method for analysing the structure of higher dimensional black objects with regular horizons is
to investigate the near-horizon limit of the black hole. This limit can be taken when the black hole is extremal,
and corresponds to a decoupling limit. In this limit, one removes information about the asymptotic behaviour of
the black hole, but retains information concerning the structure of the horizon. In addition, if the solution is
supersymmetric, supersymmetry
imposes further conditions on the near-horizon geometry. The first such classification of
supersymmetric near-horizon geometries in five dimensions was undertaken for the minimal ungauged
five-dimensional supergravity in \cite{reallbh}. Later work in \cite{adsbh}
partially generalized this analysis to the minimal gauged supergravity with negative cosmological constant.
However, in this case, even with supersymmetry, it was not possible to completely classify the near-horizon geometries,
although the conditions obtained were used to find new solutions, later generalized in \cite{pope}.
Following on from this, in \cite{kunduri1}, supersymmetric near-horizon geometries in the same theory were
examined in the special case for which the solution admits two commuting rotational Killing vectors. With this
extra assumption, it was shown that all supersymmetric black ring horizons admit conical singularities.
It is not known if this remains true when one drops the assumption that there are two commuting rotational
Killing vectors. 

This issue is closely linked to the supersymmetry preserved by black hole solutions.
In particular, all known supersymmetric black holes with a negative cosmological constant preserve
two supersymmetries. In the near-horizon limit, there is supersymmetry enhancement, and the
solutions preserve four supersymmetries. In some theories, such as minimal ungauged five-dimensional
supergravity \cite{reallbh}, and also in ten-dimensional heterotic supergravity \cite{heterotic}, the conditions imposed by
the near-horizon geometry are sufficient to ensure that the number of supersymmetries one might expect to
be preserved is automatically doubled. For the ungauged five-dimensional theory, any supersymmetric solution
preserves four supersymmetries, which is automatically doubled to eight (i.e. maximal supersymmetry)
in the near horizon limit. Also, in the heterotic theory, the conditions imposed by the near-horizon limit imply supersymmetry
doubling from $N=1$ to $N=2$. The presence of enhanced supersymmetry in turn corresponds to additional
symmetries in the solution. However, for five-dimensional black holes with negative cosmological constant, there
does not appear to be any automatic supersymmetry enhancement, and the conditions on
the geometry seem weaker.

The purpose of this paper is to classify the near-horizon geometries of ``fake" five-dimensional de-Sitter
supergravity theory coupled to vector multiplets. We shall in particular investigate so-called {\it pseudo-supersymmetric}
solutions, which admit a Killing spinor satisfying a set of fake Killing spinor equations. The integrability conditions of
these equations are sufficient to ensure that many of the bosonic field equations hold automatically,
and so we shall primarily regard pseudo-supersymmetry as a useful solution generating technique.
Attempts have also been made to construct asymptotically de-Sitter black rings, but the only known solution
exhibits a singularity  at the horizon \cite{chu}, and so the status of these solutions is also uncertain.
There are a number of reasons why one might nevertheless expect there to be a regular black ring
in the de-Sitter theory, in contrast to the anti-de-Sitter theory. Firstly, the cosmological expansion may
act to help stabilize the ring against collapse. There is also a sense in which the solutions of the
de-Sitter theory are more similar to those of the ungauged theory than those of the anti-de-Sitter theory are.
In particular, in the de-Sitter case, all solutions preserve four (fake) supersymmetries. Moreover,
there are straightforward generalizations of the BMPV black hole \cite{bmpv}  to the de-Sitter theory 
\cite{london, sabra1, sabra2, cvetic}, including multi-centred solutions \cite{sabra3}. There are no multi-centred black hole 
solutions in the anti-de-Sitter theory.

The pseudo-supersymmetric near-horizon solutions of the minimal de-Sitter supergravity theory have already
been classified in \cite{jaihz}, where it was found that the only solution is the near-horizon geometry of
the de-Sitter BMPV solution. In this paper, we generalize this analysis to the case of de-Sitter supergravity
coupled to an arbitrary number of abelian vector multiplets. We remark that the analysis of near-horizon geometries
in the de-Sitter theory is somewhat different from that in the ungauged theory.
This is because the analysis in \cite{reallbh} was able to make direct use of an earlier classification
of all supersymmetric solutions of the ungauged theory constructed in \cite{class1}, as
in this theory, the 1-form Killing spinor bilinear generated from the Killing spinor corresponds to the
timelike Killing vector field for which the event horizon is a Killing horizon. 
In contrast, in the de-Sitter theory, one does not obtain a Killing vector from the 1-form Killing spinor bilinear.
It follows that the classifications of de-Sitter supergravity solutions constructed in \cite{herdeiro1,
herdeiro2, gsb1} are not particularly well adapted for analysing the near-horizon geometries, because
the co-ordinate systems are different. Hence, to analyse the near-horizon solutions,
we revisit the analysis of the Killing spinor equation, using a basis specially adapted
to the Gaussian null co-ordinates which are used to describe the near-horizon solution.
This is simplified considerably because a large class of the solutions reduce to near-horizon solutions of
the minimal theory, which have already been classified in \cite{jaihz}.

We shall show that the pseudo-supersymmetric near-horizon solutions split into three types.
The first type corresponds to the near-horizon geometry of the de-Sitter BMPV solution found in
\cite{sabra1}. The remaining two types cannot be embedded in the minimal theory, and correspond
to horizon cross-sections which are $T^3$, with spacetime geometry $AdS_2 \times T^3$,
and to solutions with horizon cross-section $S^1 \times S^2$. This latter solution has a natural interpretation
as the near-horizon geometry of a black ring. It is therefore natural to ask if this near-horizon geometry
actually corresponds to a black ring, and it is curious that if such an object were to exist, it would
have to be a solution only of the non-minimal theory. Although the asymptotically flat supersymmetric black
ring appears in both minimal and non-minimal theories, there is no a priori reason why the same thing
should occur in the de-Sitter theory. Furthermore, the solution is entirely regular, and does not suffer from
the types of conical singularities found in the anti-de-Sitter case.

The plan of this paper is as follows. In section 2, we summarize the properties of the de-Sitter supergravity
theory, and the Gaussian null co-ordinates which are used to describe the near-horizon geometry.
In sections 3,  4 and 5 we examine the conditions on the scalars, fluxes and geometry imposed by
the existence of a non-zero (fake) Killing spinor solution of the Killing spinor equations.
In section 6 we summarize our results and present our conclusions. In Appendix A, the components of the
spin connection and curvature are presented, together with a number of conventions associated with 
the spinors. In Appendix B, we derive the co-ordinate transformations used to write the de-Sitter BMPV solution
found in \cite{sabra1} in Gaussian null co-ordinates, and obtain the near-horizon limit solution.

\section{De Sitter Supergravity in Five Dimensions}

The model we consider is $N=2$, $D=5$
gauged supergravity coupled to abelian vector multiplets \cite{gunaydin}
whose bosonic action is given by 
\begin{equation}
S={\frac{1}{16\pi G}}\int \left( R+2g^{2}{\mathcal{V}}\right) 
{\mathcal{\ast }}1-Q_{IJ}\left( dX^{I}\wedge \star dX^{J}+F^{I}\wedge \ast
F^{J}\right) -{\frac{C_{IJK}}{6}}F^{I}\wedge F^{J}\wedge A^{K}
\label{action}
\end{equation}
where $I,J,K$ take values $1,\ldots ,n$,  and $F^{I}=dA^{I}$ are the two-forms
representing gauge field strengths (one of the gauge fields corresponds to
the graviphoton). The constants $C_{IJK}$ are symmetric in $IJK$, we will
assume that $Q_{IJ}$ is invertible, with inverse $Q^{IJ}$. The $X^{I}$ are
scalar fields subject to the constraint 
\begin{equation}
X_{I}X^{I}=1 
 \label{conda}
\end{equation}
where
\begin{eqnarray}
X_I = {1 \over 6}C_{IJK} X^J X^K \ .
\end{eqnarray}

The fields $X^{I}$ can thus be regarded as being functions of $n-1$
unconstrained scalars $\phi ^{r}$. The gauge field coupling is
\begin{eqnarray}
Q_{IJ} ={\frac{9}{2}}X_{I}X_{J}-{\frac{1}{2}}C_{IJK}X^{K} 
\end{eqnarray}
which is assumed to be positive definite,
and the scalars satisfy
\begin{eqnarray}
Q_{IJ}X^{J} ={\frac{3}{2}}X_{I}\,,\qquad Q_{IJ}dX^{J}=
-{\frac{3}{2}}dX_{I} \ .
\end{eqnarray}
The scalar potential is
\begin{eqnarray}
{\mathcal{V}} &=&9V_{I}V_{J}(X^{I}X^{J}-{\frac{1}{2}}Q^{IJ}) \ ,
\end{eqnarray}
where $V_I$ are constants, which for the gauged theory do not all vanish.

Fake supergravity theory is obtained by sending $g^{2}$ to $-g^{2}$ in the
above action. The fake gravitino Killing spinor equation is given by:
\begin{equation}
\left[ \nabla _{M}-{\frac{i}{8}}\Gamma _{M}H_{N_{1}N_{2}}\Gamma
^{N_{1}N_{2}}+{\frac{3i}{4}}H_{M}{}^{N}\Gamma _{N}-
g(\frac{i}{2}X\Gamma _{M}-\frac{3}{2}A{}_{M})\right] \epsilon =0,  \label{grav}
\end{equation}
where we have defined 
\begin{equation}
\label{conv1}
V_{I}X^{I}=X,\text{ \ \ \ \ \ }V_{I}A^{I}{}_{M}=A{}_{M},\text{ \ \ \ \ \ \ }
X_{I}F^{I}{}_{MN}=H_{MN} \ ,
\end{equation}
and $\epsilon$ is a Dirac spinor. The dilatino Killing spinor equation is given by 
\begin{equation}
\label{dil}
\left( (F_{MN}^{I}-X^{I}H_{MN})\Gamma ^{MN}-2i\nabla _{M}X^{I}\Gamma
^{M}-4gV_{J}(X^{I}X^{J}-{\frac{3}{2}}Q^{IJ})\right) \epsilon =0 \ .
\end{equation}
Here we work with a metric of mostly plus signature $(-,+,+,+,+)$. The gauge field equations and the Einstein field equations
are:
\begin{eqnarray}
\label{gauge}
d \star (Q_{IJ} F^J) +{1 \over 4} C_{IJK} F^J \wedge F^K =0 \ ,
\end{eqnarray}
and
\begin{eqnarray}
\label{einst}
R_{MN}- Q_{IJ} F^I_{ML}F^J_N{}^L - Q_{IJ} \nabla_M X^I \nabla_N X^J
\nonumber \\
+ g_{MN} \bigg({1 \over 6} Q_{IJ} F^I_{L_1 L_2} F^{J \ L_1 L_2}
+6g^2 \big( {1 \over 2} Q^{IJ}- X^I X^J \big) V_I V_J \bigg) =0 \ .
\end{eqnarray}

\subsection{Gaussian Null Co-ordinates}

In order to investigate the near-horizon geometries of extremal black holes, one
first introduces Gaussian null co-ordinates adapted to the event horizon of the
black hole. It has been shown \cite{symm3} that for extremal black hole solutions
of various higher-dimensional supergravity theories satisfying certain conditions, an
event horizon of a rotating extremal black hole must be a Killing horizon, and
the solution admits a rotational isometry. However, here we shall simply
assume that the event horizon is a Killing horizon, associated
with a timelike Killing vector ${\partial \over \partial u}$, which becomes null on the horizon.
The Killing vector ${\partial \over \partial u}$ does not correspond to a Killing spinor bilinear.
We also  restrict our analysis to the case
of extremal black holes.

In this case, Gaussian null co-ordinates $u, r, y^M$ for $M=1,2,3$ adapted to
${\partial \over \partial u}$ can be introduced following the reasoning set out in \cite{gnull}. The metric is
 \begin{eqnarray}
 \label{gnullmet}
 ds^2 = -r^2 \Delta du^2 +2 du dr + 2r h du + ds_{\cal{H}}^2
 \end{eqnarray}
where the horizon is at $r=0$. $ds_{\cal{H}}^2= \gamma_{MN} dy^M dy^N$ is
the metric on spatial cross-sections of the horizon, ${\cal{H}}$, (to be distinguished from the 2-form $H$ introduced in ({\ref{conv1}})), which is analytic in
$r$ and independent of $u$, and is regular at $r=0$. $h$ is a 1-form on ${\cal{H}}$,
and $\Delta$ is a scalar on ${\cal{H}}$, which are again analytic in $r$, and independent of $u$.
As the black hole is extremal, one can take the near-horizon limit by setting
\begin{eqnarray}
r= \epsilon {\tilde{r}}, \qquad u = \epsilon^{-1} {\tilde{u}}
\end{eqnarray}
and taking $\epsilon \rightarrow 0$. On dropping the tilde on $r, u$, the near horizon
metric is of the same form as ({\ref{gnullmet}}), but with $h, \Delta, \gamma_{MN}$ independent
of both $r$ and $u$. We  assume that the spatial cross-section of the
horizon, ${\cal{H}}$,  equipped with metric $ds_{\cal{H}}^2$, is compact and simply connected.
We also assume that the resulting near-horizon geometry is pseudo-supersymmetric, though
we do not assume that the black hole bulk geometry is pseudo-supersymmetric.

It is convenient to use the following null basis adapted to the
Gaussian null co-ordinates:
\begin{eqnarray}
\label{gnullbasis}
{\bf{e}}^+ &=& du \ ,
\nonumber \\ 
{\bf{e}}^- &=& dr +rh -{1 \over 2} r^2 \Delta du \ ,
\nonumber \\ 
{\bf{e}}^i&=& e^i{}_M dy^M \ ,
 \end{eqnarray}
 where the metric is written as
 \begin{eqnarray}
 ds^2 = 2 {\bf{e}}^+ {\bf{e}}^- + \delta_{ij} {\bf{e}}^i {\bf{e}}^j \ ,
 \end{eqnarray}
 where $h=h_i(y) {\bf{e}}^i$, $\Delta=\Delta(y)$ and $e^i{}_M= e^i{}_M (y)$ depend only on 
 the co-ordinates $y$ of ${\cal{H}}$.  
 The components of the spin connection, and some other conventions associated with this basis,
are listed in the Appendix.

We further assume that the components of the gauge potentials $A^I$ remain regular
in the near-horizon limit, and therefore take
\begin{eqnarray}
A^I_+ &=& r \Phi^I \ ,
\nonumber \\
A^I_- &=& 0 \ ,
\nonumber \\
A^I_m &=& B^I_m \ ,
\end{eqnarray}
where $\Phi^I$ and $B^I_m$ do not depend on $u, r$.
It follows that the components of the field strength are given by
\begin{eqnarray}
F^I_{+-} &=& - \Phi^I \ ,
\nonumber \\
F^I_{-m} &=& 0 \ ,
\nonumber \\
F^I_{+m} &=& r \big(-\partial_m \Phi^I + \Phi^I h_m\big) \ ,
\nonumber \\
F^I_{mn} &=& (dB^I)_{mn} \ .
\end{eqnarray}
It will also be convenient to define the two form ${\hat{H}}$ as
\begin{eqnarray}
{\hat{H}}= {1 \over 2} H_{ij} {\bf{e}}^i \wedge {\bf{e}}^j \ .
\end{eqnarray}
In addition, we assume that the scalars $X^I$ also remain regular in the near-horizon limit, so 
$X^I = X^I (y)$ are smooth functions on ${\cal{H}}$.

\section{Analysis of the Killing spinor equations}

In this section, we shall begin to analyse the Killing spinor equations ({\ref{grav}}) and ({\ref{dil}}),
concentrating on the $M=-$ component of ({\ref{grav}}), and ({\ref{dil}}). The remaining components of
({\ref{grav}}) will be analysed in the following sections.
To begin, we consider the $M=-$ component of ({\ref{grav}}). On decomposing the
Dirac Killing spinor $\epsilon=\epsilon_+ + \epsilon_-$ into positive and negative chirality parts, as described
in the Appendix, one finds that this component of the gravitino equation is:
\begin{eqnarray}
\partial_r \epsilon_+ &=& 0 \ ,
\nonumber \\
\partial_r \epsilon_- &=& \Gamma_- \big({1 \over 4} h_i \Gamma^i +{i \over 2} H_{+-}
+{i \over 2} gX +{i \over 8} H_{ij} \Gamma^{ij} \big) \epsilon_+ \ ,
\end{eqnarray}
and hence we find
\begin{eqnarray}
\label{sol1}
\epsilon_+ &=& \phi_+ \ ,
\nonumber \\ 
\epsilon_- &=& r \Gamma_- \big( {1 \over 4}(h+ \star_3 {\hat{H}})_i \Gamma^i +{i \over 2}(H_{+-}+gX) \big) \phi_+ + \phi_- \ ,
\end{eqnarray}
where $\Gamma_{\pm} \phi_\pm =0$ and $\phi_{\pm}$ do not depend on $r$.

Next, consider the dilatino equation ({\ref{dil}}). On decomposing this into positive and negative chirality parts,
one obtains
\begin{eqnarray}
\label{pos1}
\bigg( -2 (F^I_{+-} - X^I H_{+-})-2i \star_3 (dB^I - X^I X_J dB^J)_i \Gamma^i
\nonumber \\
-2i \partial_i X^I \Gamma^i -4g V_I(X^I X^J-{3 \over 2}Q^{IJ}) \bigg) \epsilon_+=0
\end{eqnarray}
and
\begin{eqnarray}
\label{pos2}
\bigg( 2 (F^I_{+-} - X^I H_{+-})+2i \star_3 (dB^I - X^I X_J dB^J)_i \Gamma^i
\nonumber \\
-2i \partial_i X^I \Gamma^i -4g V_I(X^I X^J-{3 \over 2}Q^{IJ}) \bigg) \epsilon_-
\nonumber \\
+2r \bigg( (-\partial_i \Phi^I + \Phi^I h_i) - X^I X_J (-\partial_i \Phi^J + \Phi^J h_i) \bigg) \Gamma_- \Gamma^i \epsilon_+=0 \ .
\end{eqnarray}

\subsection{Solutions with both $\phi_+ \neq 0$ and $\phi_- \neq 0$}

We shall consider first the case for which both $\phi_+ \neq 0$ and $\phi_- \neq 0$.
On substituting ({\ref{sol1}) into ({\ref{pos1}}) and ({\ref{pos2}}) and examining the
$r$-independent terms in the resulting equations, one finds that
\begin{eqnarray}
V_J (X^I X^J -{3 \over 2} Q^{IJ}) =0
\end{eqnarray}
or equivalently
\begin{eqnarray}
X X_I = V_I \ .
\end{eqnarray}
Note that as we are interested in solutions of the {\it gauged} theory, for which not all $V_I$ vanish,
this condition implies that $X \neq 0$. Furthermore, one finds that
\begin{eqnarray}
\label{aux1a}
dX^I=0, \qquad F^I_{+-}= X^I H_{+-}, \qquad dB^I = X^I X_J dB^J \ .
\end{eqnarray}
Also from the term linear in $r$ in equation ({\ref{pos2}}) one finds that
\begin{eqnarray}
F^I_{+i} = X^I H_{+i}
\end{eqnarray}
which together with ({\ref{aux1a}}) implies that
\begin{eqnarray}
F^I = X^I H \ .
\end{eqnarray}
Taken together, all of these conditions imply that the solution reduces to a solution of the
{\it minimal} de-Sitter supergravity theory in five dimensions. However, all pseudo-supersymmetric solutions
of this theory have been fully classified in \cite{jaihz}, and in this case, the near horizon geometries
are all isometric to the near horizon geometry of the pseudo-supersymmetric extremal de-Sitter BMPV solution
found in \cite{london,  sabra2}. We shall therefore henceforth consider only the remaining two cases, for which
either $\phi_+ =0, \phi_- \neq 0$, or $\phi_- =0, \phi_+ \neq0$.

\subsection{Solutions with either $\phi_+=0, \phi_- \neq 0$, or $\phi_- =0, \phi_+ \neq 0$}

For these remaining cases, we first note that the dilatino Killing spinor equation ({\ref{pos1}}) and
({\ref{pos2}}) implies that
\begin{eqnarray}
d X^I = \pm \star_3 (dB^I - X^I X_J dB^J) \ .
\end{eqnarray}
On taking the divergence of this equation, one finds
\begin{eqnarray}
{\hat{\nabla}}^i {\hat{\nabla}}_i X^I = \mp {\hat{\nabla}}^i X^I \star_3 (X_J dB^J)_i \mp X^I {\hat{\nabla}}^i (\star_3 (X_J dB^J)_i )
\end{eqnarray}
and hence
\begin{eqnarray}
X_I {\hat{\nabla}}^i {\hat{\nabla}}_i X^I =  \mp {\hat{\nabla}}^i (\star_3 (X_J dB^J)_i ) \ .
\end{eqnarray}
It follows that
\begin{eqnarray}
\int_{\cal{H}} X_I {\hat{\nabla}}^i {\hat{\nabla}}_i X^I = 0 \ .
\end{eqnarray}
This condition can be rewritten, on partially integrating, as
\begin{eqnarray}
\int_{\cal{H}} Q_{IJ} {\hat{\nabla}}_i X^I {\hat{\nabla}}^i X^J =0
\end{eqnarray}
and assuming that $Q_{IJ}$ is positive definite, this in turn implies that
the scalars $X^I$ are constant, and moreover
\begin{eqnarray}
dB^I = X^I {\hat{H}} \ .
\end{eqnarray}

In the following two sections, we shall consider the two cases $\phi_+=0, \phi_- \neq 0$ and $\phi_- =0, \phi_+ \neq 0$
separately, and in greater detail.

\section{Further Analysis: Solutions with $\phi_+ \neq 0, \phi_- =0$}

For solutions with $\phi_+ \neq 0, \phi_- =0$, note that ({\ref{pos1}}) and ({\ref{pos2}}) imply that
\begin{eqnarray}
\label{van1}
V_J (X^I X^J -{3 \over 2} Q^{IJ}) (H_{+-}+gX)=0 \ .
\end{eqnarray}
It is straightforward to show that if $V_J (X^I X^J -{3 \over 2} Q^{IJ})=0$ for all $I$, then ({\ref{pos1}})
and ({\ref{pos2}}) also imply that $F^I = X^I H$. Hence the solution reduces to one of the minimal theory,
and as mentioned previously, such solutions have been fully classified in \cite{jaihz}, and so we discard this case.
Therefore we obtain from ({\ref{van1}})
\begin{eqnarray}
H_{+-}=-gX \ .
\end{eqnarray}
The remaining content of ({\ref{pos1}}) and ({\ref{pos2}}) implies that
\begin{eqnarray}
\Phi^I = 3g(X X^I - Q^{IJ} V_J)
\end{eqnarray}
so, in particular, $\Phi^I$ are constant, as well as
\begin{eqnarray}
{\hat{H}} = \star_3 h
\end{eqnarray}
and hence one finds
\begin{eqnarray}
d \star_3 h =0 \ .
\end{eqnarray}
With these conditions, one finds that the Killing spinor can be simplified to
\begin{eqnarray}
\epsilon = \phi_+ + {1 \over 2} r \Gamma_- h_i \Gamma^i \phi_+
\end{eqnarray}
which can be used to simplify the remaining components of ({\ref{grav}}).

Next, consider the $M=+$ component of ({\ref{grav}}). From the positive chirality component of this equation, one finds
\begin{eqnarray}
\partial_u \phi_+ =0
\end{eqnarray}
so that $\phi_+$ depends only on the $y$ co-ordinates. The positive chirality component of this equation also implies that
\begin{eqnarray}
\Delta = -3g V_I \Phi^I
\end{eqnarray}
so $\Delta$ is constant, as well as
\begin{eqnarray}
\label{dh1}
dh = 3gX \star_3 h \ .
\end{eqnarray}
Given these conditions, the negative chirality component of the $M=+$ component of ({\ref{grav}}) vanishes.

It remains to consider the $M=i$ component of ({\ref{grav}}). From the positive chirality component of this, one finds
\begin{eqnarray}
\label{cd1}
{\hat{\nabla}}_i \phi_+ + \bigg(-{3i \over 4} gX \Gamma_i +{i \over 2} {\hat{H}}_{ij} \Gamma^j +{3 \over 2}g A_i
-{1 \over 2} h_i \bigg) \phi_+ =0
\end{eqnarray}
and the negative component of  this component of ({\ref{grav}}) is equivalent to
\begin{eqnarray}
{\hat{\nabla}}_{(i} h_{j)} =0
\end{eqnarray}
so $h$ defines a Killing vector on ${\cal{H}}$. Note that this condition, together with ({\ref{dh1}}) implies that
$h^2$ is constant. Lastly, we compute the integrability condition associated with ({\ref{cd1}}) in order to obtain the following 
expression for the Ricci tensor ${\hat{R}}_{ij}$ of ${\cal{H}}$;
\begin{eqnarray}
{\hat{R}}_{ij} =  \bigg( h^2 +{9 \over 2} g^2 X^2 \bigg) \delta_{ij} - h_i h_j \ .
\end{eqnarray}

\section{Further Analysis: Solutions with $\phi_- \neq 0, \phi_+ =0$}

If $\phi_- \neq 0, \phi_+ =0$, then ({\ref{pos1}}) and ({\ref{pos2}}) simplify to
\begin{eqnarray}
\label{auxx1}
-\Phi^I + X^I X_J \Phi^J -2g V_J (X^I X^J -{3 \over 2} Q^{IJ} ) =0
\end{eqnarray}
and the positive and negative chirality parts of the $M=+$ component of ({\ref{grav}}) are
\begin{eqnarray}
\label{auxx2}
{\hat{H}} = \star_3 h, \qquad H_{+-}= gX
\end{eqnarray}
and 
\begin{eqnarray}
\label{auxx3}
\partial_u \phi_- =0, \qquad \Delta = 3g V_I \Phi^I, \qquad -\star_3 dh +3 X_I (- \Phi^I + \Phi^I h) =0
\end{eqnarray}
respectively.
Using ({\ref{auxx2}}), one can simplify ({\ref{auxx1}}) to give
\begin{eqnarray}
\Phi^I = -3g (X X^I - Q^{IJ} V_J)
\end{eqnarray}
so, in particular, $d \Phi^I=0$, and hence ({\ref{auxx3}}) can be simplified further to give
\begin{eqnarray}
\label{dh2}
dh = - 3 g X \star_3 h \ .
\end{eqnarray}
Furthermore, ({\ref{auxx2}}) implies that
\begin{eqnarray}
d \star_3 h =0 \ .
\end{eqnarray}
Next, consider the $M=i$ component of ({\ref{grav}}); this has no positive chirality component, and the negative chirality
component is
\begin{eqnarray}
\label{cd2}
{\hat{\nabla}}_i \phi_- + \bigg(-{3i \over 4} gX \Gamma_i +{i \over 2} {\hat{H}}_{ij} \Gamma^j +{3 \over 2}g A_i
+{1 \over 2} h_i \bigg) \phi_- =0 \ .
\end{eqnarray}
After some computation, one finds from the integrability condition of ({\ref{cd2}}) the following expression for the Ricci tensor of
${\cal{H}}$:
\begin{eqnarray}
\label{rtens2}
{\hat{R}}_{ij} =  -2 {\hat{\nabla}}_{(i} h_{j)} + \bigg( h^2 +{9 \over 2} g^2 X^2 \bigg) \delta_{ij} - h_i h_j \ .
\end{eqnarray}
To proceed, define 
\begin{eqnarray}
I = \int_{\cal{H}} 2 {\hat{\nabla}}_{(i} h_{j)} {\hat{\nabla}}^{(i} h^{j)}
\end{eqnarray}
and note that one can rewrite 
\begin{eqnarray}
I = {1 \over 2} \int_{\cal{H}} (dh)_{ij} (dh)^{ij} + 2  \int_{\cal{H}} {\hat{\nabla}}_j h_i {\hat{\nabla}}^i h^j \ .
\end{eqnarray}
On using ({\ref{dh2}}) to simplify the first integral, and partially integrating the second integral, one finds
\begin{eqnarray}
I =  \int_{\cal{H}} 9 g^2 X^2 h^2 -2 h^i {\hat{\nabla}}_j {\hat{\nabla}}_i h^j  = \int_{\cal{H}} 9 g^2 X^2 h^2 -2 h^i ({\hat{\nabla}}_j {\hat{\nabla}}_i - {\hat{\nabla}}_i {\hat{\nabla}}_j) h^j
\end{eqnarray}
where we have made use of $d \star_3 h=0$ to further simplify the second integral.
It follows that
\begin{eqnarray}
I =  \int_{\cal{H}} 9 g^2 X^2 h^2 -2 h^i h^j R_{ij}
\end{eqnarray}
and on substituting the Ricci tensor given in ({\ref{rtens2}}) into this equation and using
\begin{eqnarray}
 \int_{\cal{H}} h^i h^j {\hat{\nabla}}_i h_j ={1 \over 2}  \int_{\cal{H}} h^i {\hat{\nabla}}_i h^2 = -{1 \over 2}  \int_{\cal{H}} h^2 {\hat{\nabla}}^i h_i =0
\end{eqnarray}
one finds that $I=0$, so 
\begin{eqnarray}
{\hat{\nabla}}_{(i} h_{j)}=0 \ .
\end{eqnarray}
This, together with ({\ref{dh2}}), implies that $h^2$ is constant. So the Ricci tensor of ${\cal{H}}$ ({\ref{rtens2}}) simplifies to
\begin{eqnarray}
{\hat{R}}_{ij} =  \bigg( h^2 +{9 \over 2} g^2 X^2 \bigg) \delta_{ij} - h_i h_j \ .
\end{eqnarray}
These conditions on the geometry and flux are therefore identical to those found for the analysis
of the solutions $\phi_+ \neq 0, \phi_- =0$, modulo a sign change in $dh$ and in the $du \wedge dr$ component of $F^I$,
however compactness of ${\cal{H}}$ has been used in a different way to find these conditions.

\section{Conclusions}

Combining the various conditions derived in the previous sections,
one finds that the scalars $X^I$ are constant, and the metric is given by
\begin{eqnarray}
ds^2 =-9 g^2 r^2(Q^{IJ} - X^I X^J)V_I V_J  du^2 +2 du dr +2r du h + ds_{\cal{H}}^2 \ .
\end{eqnarray}
The metric $ds_{\cal{H}}^2$ on the spatial cross-sections of the horizon has
Ricci tensor
\begin{eqnarray}
\label{rsc3}
{\hat{R}}_{ij} =  \bigg( h^2 +{9 \over 2} g^2 (V_I X^I)^2  \bigg) \delta_{ij} - h_i h_j \ .
\end{eqnarray}
Moreover, the 1-form $h$ on ${\cal{H}}$ is Killing, ${\hat{\nabla}}_{(i} h_{j)}=0$, with $h^2$ constant, and satisfies
\begin{eqnarray}
dh = \pm 3g V_I X^I \star_3 h
\end{eqnarray}
and the gauge field strengths are 
\begin{eqnarray}
F^I = \mp 3g (X^I X^J - Q^{IJ}) V_J du \wedge dr + X^I \star_3 h \ .
\end{eqnarray}
We remark that this solution satisfies the gauge and Einstein field equations ({\ref{gauge}}), ({\ref{einst}}).
As a consequence of the integrability conditions of the Killing spinor equations, the scalar field equations are also
automatically satisfied.

On examining the Ricci tensor given in ({\ref{rsc3}}), and using the argument given in \cite{reallbh}, one finds a number of possible geometries for ${\cal{H}}$:
\begin{itemize}
\item[(i)] If $V_I X^I \neq 0$ then ${\cal{H}}$ is a squashed $S^3$ if $h \neq 0$, and a round $S^3$ if $h=0$.
\item[(ii)] If $V_I X^I=0$ and $h=0$ then ${\cal{H}}$ is $T^3$.
\item[(iii)] If $V_I X^I=0$ and $h \neq 0$ then ${\cal{H}}$ is $S^1 \times S^2$.
\end{itemize}

We remark that the solution for which $V_I X^I \neq 0$ corresponds to the near-horizon 
geometry of the extremal de-Sitter BMPV solution  \cite{sabra1}. The near-horizon geometry of
this solution is derived in Appendix B. The solution for which $V_I X^I=0$ and $h=0$ has spacetime geometry
$AdS_2 \times T^3$.

A special subclass of solutions with $V_I X^I \neq 0$ corresponds
to the near horizon solutions of the minimal theory. This case corresponds to taking
\begin{eqnarray}
X^1= \sqrt{3}, \qquad X_1 = {1 \over \sqrt{3}}, \qquad C_{111} = {2 \over \sqrt{3}}, \qquad Q_{11}={1 \over 2}
\end{eqnarray}
and the resulting solution, on setting $\ell^2 = {1 \over 3 (g V_1)^2}$, corresponds 
to the near-horizon geometry of the extremal pseudo-supersymmetric BMPV 
de-Sitter black hole solution, as obtained in \cite{jaihz}.
For solutions with $V_I X^I \neq 0$, in the special case when $h=0$, the spacetime geometry
is $M_2 \times S^3$ where the metric on the 2-manifold $M_2$ is 
\begin{eqnarray}
ds_2^2 = -r^2 \Delta du^2 +2 du dr, \qquad \Delta = 9g^2 (Q^{IJ}-X^I X^J) V_I V_J
\end{eqnarray}
and so $M_2$ is $AdS_2$, $dS_2$ or ${\mathbb{R}}^{1,1}$ according as $\Delta >0$, $\Delta<0$ or $\Delta=0$.
For fixed scalars $X^I$ satisfying ({\ref{conda}}), in the non-minimal theory, one can always choose $V_I$ such that both $V_I X^I \neq 0$ and $\Delta >0$, or $\Delta <0$, or $\Delta =0$, i.e. $M_2$ is $AdS_2$, $dS_2$ or ${\mathbb{R}}^{1,1}$. In comparison,
in the minimal theory, $\Delta$ is always negative, so $M_2$ is $dS_2$.

The solution with $V_I X^I =0$ but $h \neq 0$ is also particularly interesting. This solution 
cannot be embedded in the minimal theory.
As the spatial cross-sections of the horizon are $S^1 \times S^2$ in this case, the solution
has a natural interpretation as the near-horizon geometry of a de-Sitter black ring. 
This does not contradict the results of \cite{jaihz}, as this analysis was restricted to the
near-horizon solutions of the minimal theory. An analogous type of black ring
near-horizon geometry was found for the non-minimal anti-de-Sitter five-dimensional
supergravity theory in \cite{kunduri2}. In that case,
the scalars were shown to be constant, though the condition they must satisfy, and
which cannot hold for the minimal theory, differs
from the condition $V_I X^I=0$ under consideration here. Furthermore, the candidate $AdS$ black ring near horizon
geometries are $AdS_3 \times S^2$, which also differs from the solutions found for the de-Sitter case.
It is currently unknown whether the black ring near-horizon geometry found for the anti-de-Sitter theory in 
\cite{kunduri2} actually corresponds to an asymptotically $AdS_5$ black ring supergravity solution.
In order to determine the spacetime geometry for the de-Sitter near-horizon solutions
with $V_I X^I=0$ and $h \neq 0$, note that as $h$ is closed, we set locally  $h= d \psi$.
One finds that the spacetime is $M_3 \times S^2$, with metric
\begin{eqnarray}
ds^2 = (k  d \psi +{r \over k} du)^2 - r^2 (\Delta +{1 \over k^2}) du^2 +2du dr
+2k^2 (d \theta^2+ \sin^2 \theta d \phi^2)
\end{eqnarray}
for nonzero constant $k$, with $\Delta = 9g^2 Q^{IJ} V_I V_J >0$, and the metric on $M_3$ is given by
\begin{eqnarray}
\label{tmet}
ds_3^2 =  (k  d \psi +{r \over k} du)^2 - r^2 (\Delta +{1 \over k^2}) du^2 +2du dr \ .
\end{eqnarray}
This metric is not Einstein, but is a $U(1)$ fibration over $AdS_2$.
Curiously, the metric ({\ref{tmet}}) is related to the near-horizon extremal Kerr (NHEK) solution.
This can be written as \cite{nhek1}
\begin{eqnarray}
\label{nhek1}
ds^2 = {1 \over 2}(1+\cos^2 \chi ) \bigg[ -{r^2 \over r_0^2} du^2 + 2 du d r
+ r_0^2 d \chi^2 \bigg] +{2 r_0^2 \sin^2 \chi \over 1+\cos^2 \chi} \bigg(d \sigma + {r \over r_0^2}
du \bigg)^2 \ .
\end{eqnarray}
Although the NHEK geometry written in this fashion is not in Gaussian Null form,
we shall concentrate on the pull-back of the metric to slices $\chi=const$ ($\sin \chi \neq 0$). After taking the pull-back,
we also set 
\begin{eqnarray}
\sigma= {(1+\cos^2 \chi) \over 4 \sin^2 \chi} \psi \ .
\end{eqnarray}
The resulting metric is then given by
\begin{eqnarray}
\label{nhek2}
{\hat{ds}}^2 = - {2\big(\cos^4 \chi +6 \cos^2 \chi -3 \big) \over r_0^2 (1+\cos^2 \chi)^3} r^2 du^2
+2 du dr + 2r d \psi du + {r_0^2 (1+\cos^2 \chi) \over 8 \sin^2 \chi} d \psi^2\ ,
\nonumber \\
\end{eqnarray}
which is in Gaussian Null form.
It is also clear that this slicing of the NHEK metric is then locally isometric to the 
metric on $M_3$ given in ({\ref{tmet}}) on making the identifications
\begin{eqnarray}
k^2 = {(1+\cos^2 \chi) r_0^2 \over 8 \sin^2 \chi}, \qquad  
\Delta = {2 \over r_0^2 (1+\cos^2 \chi)^3} \bigg(\cos^4 \chi +6 \cos^2 \chi -3 \bigg)
\end{eqnarray}
and imposing the restriction $0 < \chi < \arcsin ({\sqrt{3}-1})$ in order to have $\Delta>0$.

It remains to determine if this candidate black ring near horizon geometry actually corresponds to
a genuine black ring solution, which is work in progress.

\section*{Acknowledgments}

The work of WS was supported in part by the National Science Foundation
under grant number PHY-0903134. JG is supported by the EPSRC grant
EP/F069774/1.

 \setcounter{section}{0}

\appendix{}

The non-vanishing components of the spin connection are given by
\begin{eqnarray}
\Omega_{+,+-} = -r \Delta, \quad \Omega_{+,+i} = {1 \over 2} r^2 (\Delta h_i - \partial_i \Delta),
\quad \Omega_{+,-i} = -{1 \over 2} h_i, \quad \Omega_{+,ij} = -{1 \over 2} r (dh)_{ij}
\nonumber \\
\end{eqnarray}
and
\begin{eqnarray}
\Omega_{-,+i} = -{1 \over 2} h_i
\end{eqnarray}
and
\begin{eqnarray}
\Omega_{i,+-} ={1 \over 2} h_i, \qquad \Omega_{i,+j} = -{1 \over 2} r (dh)_{ij},
\qquad \Omega_{ijk} = \omega_{ijk}
\end{eqnarray}
where $\omega_{ijk}$ is the spin connection of the horizon section ${\cal{H}}$.
We denote the Levi-Civita connection of ${\cal{H}}$ by ${\hat{\nabla}}$. We note that the Ricci tensor
of the 5-dimensional near-horizon geometry has components
\begin{eqnarray}
R_{+-} &=& {1 \over 2} {\hat{\nabla}}^i h_i -\Delta -{1 \over 2}h^2 \ ,
\nonumber \\
R_{ij} &=& {\hat{R}}_{ij} + {\hat{\nabla}}_{(i} h_{j)} -{1 \over 2} h_i h_j \ ,
\nonumber \\
R_{++} &=& r^2 \big( {1 \over 2} {\hat{\nabla}}^2 \Delta -{3 \over 2} h^i {\hat{\nabla}}_i \Delta -{1 \over 2} \Delta {\hat{\nabla}}^i h_i 
+ \Delta h^2 +{1 \over 4} (dh)_{ij} (dh)^{ij} \big) \ ,
\nonumber \\
R_{+i} &=&r \big( {1 \over 2} {\hat{\nabla}}^j (dh)_{ij} - (dh)_{ij} h^j - {\hat{\nabla}}_i \Delta + \Delta h_i \big) \ ,
\nonumber \\
R_{--} &=& 0 \ ,
\nonumber \\
R_{-i} &=&0 \ ,
\end{eqnarray}
where ${\hat{R}}_{ij}$ denotes the Ricci tensor of ${\cal{H}}$.

We shall decompose Dirac spinors $\epsilon$ as
\begin{eqnarray}
\epsilon = \epsilon_+ + \epsilon_-
\end{eqnarray}
where
\begin{eqnarray}
\Gamma_\pm  \epsilon_\pm =0 \ ,
\end{eqnarray}
and we adopt the null basis given in ({\ref{gnullbasis}}). In particular, note that
\begin{eqnarray}
\Gamma_{+-} \epsilon_\pm = \pm \epsilon_\pm
\end{eqnarray}
and so, using an abuse of terminology, we refer to spinors which are annihilated by
$\Gamma_+$ and $\Gamma_-$ as being of positive and negative chirality respectively.

Our duality conventions are as follows; the 5-dimensional volume form $\epsilon^{(5)}$ satisfies
\begin{eqnarray}
\Gamma_{N_1 N_2 N_3 N_4 N_5} = -i \epsilon^{(5)}_{N_1 N_2 N_3 N_4 N_5}
\end{eqnarray}
and we take the volume form of the spatial cross-sections of the horizon ${\cal{H}}$, $\epsilon_{\cal{H}}$ to be given by
\begin{eqnarray}
\epsilon^{(5)}= {\bf{e}}^+ \wedge {\bf{e}}^- \wedge \epsilon_{\cal{H}} \ .
\end{eqnarray}
We denote by $\star_3$ the Hodge dual on ${\cal{H}}$ taken with respect to $\epsilon_{\cal{H}}$, and also note that
\begin{eqnarray}
\Gamma_{ijk} \epsilon _\pm &=& \mp i \epsilon_{ijk} \epsilon_{\pm}
\nonumber \\
\Gamma_{ij} \epsilon_\pm &=& \mp i \epsilon_{ij}{}^k \Gamma_k \epsilon_\pm
\end{eqnarray}
where we have dropped the subscript ${\cal{H}}$ on $\epsilon_{\cal{H}}$.
Finally, we remark that our analysis of a number of algebraic conditions obtained from the Killing spinor equations
proceeds by noting that a condition of the form
\begin{eqnarray}
(\alpha + i \beta_i \Gamma^i) \phi =0
\end{eqnarray}
for real $\alpha$, $\beta_i$ and $\phi \neq 0$, is equivalent to $\alpha=0$, $\beta=0$.

\appendix{}

In this appendix, we derive the near-horizon geometry, including the scalars and the gauge field strengths,
for the de-Sitter BMPV solution found in \cite{sabra1}. To begin, we note that all pseudo-supersymmetric solutions 
for which the 1-form Killing spinor bilinear is timelike have been classified in \cite{gsb1}.
Using the formalism in that paper, the de-Sitter BMPV solution can be written as
\begin{eqnarray}
ds^2 &=& -G^{-4} \big(dt+{q \over \rho^2} \sigma^3 \big)^2 + G^2
\big[d \rho^2 +{1 \over 4} \rho^2 ((\sigma^1)^2+(\sigma^2)^2+(\sigma^3)^2) \big] \ ,
\nonumber \\
F^I &=& d \big( G^{-2} X^I (dt +{q \over \rho^2} \sigma^3) \big) \ ,
\nonumber \\
X_I &=& G^{-2} \big(2gt V_I +{k_I \over \rho^2} \big) \ ,
\end{eqnarray}
where the function $G$ depends on $t$ and $\rho$ and is fixed by the expression for $X_I$,
and
$q, k_I$ are constants
\footnote{Using the notation of \cite{gsb1}, this solution lies within the class of solutions for which
the HKT base is conformally hyper-K\"ahler, in fact in this case the base is ${\mathbb{R}}^4$, with
${\cal{P}}=0, {\cal{Q}}={q \over \rho^2} \sigma^3, \Psi^I=0, Z_I={k_I \over \rho^2}$, and we have
re-labelled the $u$ co-ordinate used in \cite{gsb1} as $t$ for convenience.}.
The invariant 1-forms on $SU(2)$ are given by
\begin{eqnarray}
\sigma^1 &=& -\sin \psi d \theta + \cos \psi \sin \theta d \phi \ ,
\nonumber \\
\sigma^2  &=& \cos \psi d \theta + \sin \psi \sin \theta d \phi \ ,
\nonumber \\
\sigma^3 &=& d \psi + \cos \theta d \phi \ .
\end{eqnarray}
The co-ordinate transformations used to write this solution in Gaussian Null co-ordinates are similar
to those used to transform the de-Sitter BMPV solution in the minimal theory \cite{jaihz}. First one sets
\begin{eqnarray}
t = e^v, \qquad \rho = e^{-{v \over 2}} R, \qquad H = e^{-{v \over 2}} G
\end{eqnarray}
so the solution becomes
\begin{eqnarray}
ds^2 &=& - H^{-4} \big(dv +{q \over R^2} \sigma^3 \big)^2
+H^2 \big[(dR-{1 \over 2} R dv)^2 +{1 \over 4}R^2 ((\sigma^1)^2+(\sigma^2)^2+(\sigma^3)^2)] \ ,
\nonumber \\
F^I &=& d \big( H^{-2} X^I (dv +{Q \over R^2} \sigma^3) \big) \ ,
\nonumber \\
X_I &=&  H^{-2} \big( 2gV_I +{k_I \over R^2} \big) \ ,
\end{eqnarray}
and observe that $H$ depends only on $R$.
Next make the following co-ordinate transformation:
\begin{eqnarray}
dv &=& du  +{2 (H^6 R^6 -4 q^2 \pm 2 R^2 \sqrt{H^6 R^6-4q^2}) 
\over R (H^6 R^6 -4 R^4 -4 q^2)} dR \ ,
\nonumber \\
d \psi &=& d \psi' + c du + {8qR  (H^6 R^6 -4 q^2 \pm 2 R^2 \sqrt{H^6 R^6-4q^2}) 
\over (H^6 R^6 -4 R^4 -4 q^2) (H^6 R^6 -4 q^2)} dR \ ,
\end{eqnarray}
where $c$ is a constant which will be fixed later.
Furthermore, it will be convenient to write
\begin{eqnarray}
f(r) = R^2 H^2
\end{eqnarray}
where $f$ satisfies $f(0)>0$, and
\begin{eqnarray}
f {\dot{f}} = \pm 4 g X \sqrt{f^3 -4 q^2}
\end{eqnarray}
and ${\dot{f}}= {df \over dr}$. This condition defines $r$ in terms of $R$, and we note the following useful relations
\begin{eqnarray}
{\dot{f}} = 2gX {d \over dr} \big( f H^{-2} \big), \qquad H^{-2} = {1 \over 2gf X} (f-  k_I X^I)
\end{eqnarray}
and
\begin{eqnarray}
{\dot{X}} = {\dot{f} \over fX} (X^2-Q^{IJ} V_I V_J) \ .
\end{eqnarray}

The scalars and the metric can then be rewritten entirely in terms of $f$, as
\begin{eqnarray}
X_I = f^{-1} X^{-1} (f - k_J X^J) V_I + f^{-1} k_I
\end{eqnarray}
and 
\begin{eqnarray}
ds^2 &=& \bigg({1 \over 4}(1+c^2)f - f^{-2}(cq +{1 \over 2gX}(f-k_I X^I))^2 \bigg) du^2 +2 du dr
\nonumber \\
&+& \bigg({1 \over 2} cf -2q f^{-2}(cq +{1 \over 2gX}(f-k_I X^I)) \bigg) du (d \psi' + \cos \theta d \phi)
\nonumber \\
&+& {1 \over 4} f \big( d \theta^2+\sin^2 \theta d \phi^2 \big) +\big({1 \over 4}f - q^2 f^{-2} \big)
(d \psi' + \cos \theta d \phi)^2 \ .
\end{eqnarray}
We require that this metric correspond to that of an extremal black hole in Gaussian null co-ordinates,
and so on expanding out the $g_{uu}$ and $g_{u \psi'}$ components in powers of $r$, and requiring that
the $O(r^0)$ and $O(r)$ components of $g_{uu}$, and the $O(r^0)$ component of $g_{u \psi'}$ vanish,
one obtains the conditions
\begin{eqnarray}
cq - {1 \over 2g X(0)} (f-k_I X^I)(0) = {c \over 4q} f(0)^3,
\quad c^2 \big( {f(0)^3 \over 4 q^2} -1 \big)=1, \quad 
X(0) = {4q \over 3cg f(0)^2} \ . \nonumber \\
\end{eqnarray}
On expanding out the remaining components of the metric in $r$, and taking the near-horizon limit, one obtains
\begin{eqnarray}
ds^2 &=& -9 g^2 (X^2-Q^{IJ} V_I V_J)(0) r^2 du^2 +2 du dr
\pm 8 q |{q \over c}| f(0)^{-3} r du (d \psi' + \cos \theta d \phi)
\nonumber \\
&+&{1 \over 4} f(0) (d \theta^2+\sin^2 \theta d \phi)^2 +{q^2 \over c^2} f(0)^{-2} (d \psi'+\cos \theta d \phi)^2
\end{eqnarray}
and in the near-horizon limit, the scalars are
\begin{eqnarray}
X_I = {2gq \over c f(0)} V_I + {1 \over f(0)} k_I \ .
\end{eqnarray}
Note that the 1-form $h$ is given by 
\begin{eqnarray}
h = \pm 4 q |{q \over c}| f(0)^{-3} (d \psi' + \cos \theta d \phi)
\end{eqnarray}
and the metric on ${\cal{H}}$ is
\begin{eqnarray}
ds_{\cal{H}}^2 = {1 \over 4} f(0) (d \theta^2+\sin^2 \theta d \phi)^2 +{q^2 \over c^2} f(0)^{-2} (d \psi'+\cos \theta d \phi)^2 \ .
\end{eqnarray}
If one takes the volume form on ${\cal{H}}$ to be
\begin{eqnarray}
\epsilon = {q \over 4c} \sin \theta d \phi \wedge d \theta \wedge d \psi'
\end{eqnarray}
and fixes the sign so that $\pm |{q \over c}| {c \over q}=1$, 
then the gauge field strengths, evaluated in the near-horizon limit, are
\begin{eqnarray}
F^I =   -3g (X X^I - Q^{IJ} V_J)(0) du \wedge dr
+ X^I(0) \star_3 h \ .
\end{eqnarray}
Further simplification can be made by setting
\begin{eqnarray}
k = 4q f(0)^{-2}
\end{eqnarray}
so that $h^2=k^2$. Then one can rewrite
\begin{eqnarray}
ds_{\cal{H}}^2 = {1 \over k^2+9g^2 X^2} (d \theta^2+\sin^2 \theta d \phi^2) +{9g^2 X^2 \over (k^2+9g^2X^2)^2} (d \psi'+
\cos \theta d \phi)^2
\end{eqnarray}
where here we set $X=X(0)$, and also
\begin{eqnarray}
h = {3kgX \over k^2+9g^2 X^2} (d \psi'+\cos \theta d \phi) \ .
\end{eqnarray}

\renewcommand{\theequation}{A-\arabic{equation}} 

\setcounter{equation}{0}

\end{document}